# Tailoring photoluminescence by strain-engineering in layered perovskite flakes

*Davide Spirito[1], María Barra-Burillo[2], Francesco Calavalle[2], Costanza Lucia Manganelli[1], Marco Gobbi[2,3,4], Rainer Hillenbrand[2,3,5], Fèlix Casanova[2,3], Luis E. Hueso[2,3], Beatriz Martín-García [2]\**

[1]IHP–Leibniz-Institut für innovative Mikroelektronik, Im Technologiepark 25, 15236 Frankfurt (Oder) (Germany)

[2]CIC nanoGUNE BRTA, Tolosa Hiribidea, 76, 20018 Donostia-San Sebastián, Basque Country (Spain)

[3]IKERBASQUE, Basque Foundation for Science, 48013 Bilbao, Basque Country, Spain

[4]Materials Physics Center CSIC-UPV/EHU, 20018 Donostia-San Sebastián, Spain

[5]Department of Electricity and Electronics, UPV/EHU, 20018 Donostia-San Sebastián, Spain

ABSTRACT. Strain is an effective strategy to modulate the optoelectronic properties of 2D materials, but it has been almost unexplored in layered hybrid organic-inorganic metal halide perovskites (HOIPs) due to their complex band structure and mechanical properties. Here, we investigate the temperature-dependent micro-photoluminescence (PL) of 2D $(C_6H_5CH_2CH_2NH_3)_2Cs_3Pb_4Br_{13}$ HOIP subject to biaxial strain induced by a $SiO_2$ ring platform on which flakes are placed by viscoelastic stamping. At 80K, we found that a strain <1% can change the PL emission from a single peak (unstrained) to three well-resolved peaks. Supported by micro-Raman spectroscopy we show that the thermo-mechanically generated-strain modulates the bandgap due to changes in the octahedral tilting and lattice expansion. Mechanical simulations demonstrate the co-existence of tensile and compressive strain along the flake. The observed PL peaks add an interesting feature to the rich phenomenology of photoluminescence in 2D HOIPs, which can be exploited in tailored sensing and optoelectronic devices.

MAIN TEXT. Strain engineering is a powerful strategy to modulate the optoelectronic properties of 2D materials for applications in nanoelectronics and optoelectronics. Strain can be generated by using flexible substrates, controlled wrinkling, thermal expansion mismatch or patterning the substrate.[1] The exploration of this technique began with graphene and other 2D materials showing that strain could affect their crystal structure,[2] modulate their bandgap,[1,3–5] or vary their phase transitions[6], and it has been progressively refined towards dynamic modulation strategies[7]. Nevertheless, strain engineering is in an early stage for hybrid organic-inorganic metal halide perovskites (HOIPs), in spite of their demonstrated extraordinary performance in photovoltaics, LEDs and photodetectors.[8–18] Efficient approaches for inducing strain in the crystal structure have been widely investigated in 3D HOIPs during the growth of single crystals[19] or thin films[20]. These studies have shown that it is possible to modulate the bandgap and achieve enhanced device performance without changing composition.[21–23] In contrast to 3D HOIPs, strain engineering of layered (2D) HOIPs, has barely been explored. However, studies on few-layer sheets grown by solution-assisted methods[24] and on flakes placed on flexible polymers[25,26] or patterned substrates[27] have shown that these materials exhibit a modulation of photoluminescence emission and carrier transport at room-temperature and established their potential use in sensors[26]. Indeed, 2D HOIPs are ideal candidates for the study of strain engineering: these materials show improved environmental stability vs their 3D counterparts,[11] structure-tunable optoelectronic properties,[28] and a dynamic crystal structure resisting up to 6% strain before fracture[29]. Moreover, detailed mechanical studies on 2D HOIP flakes[25,29–32] have shown that their mechanical properties depend on the number of octahedral sheets per inorganic layer (denoted as *n*) and the composition (halide and organic cation), enabling to tune the material's properties. However, more efforts are needed to successfully control, design, and exploit the strain−bandgap modulation in optoelectronic and sensing devices. Here, we study the strain-response of $(C_6H_5CH_2CH_2NH_3)_2(Cs)_3Pb_4Br_{13}$ flakes placed on $SiO_2$ rings by temperature-dependent micro-photoluminescence (PL) and Raman spectroscopy. In contrast to previous reports focusing on lead iodide 2D HOIPs combined with alkyl cations,[25–27] we have selected $(C_6H_5CH_2CH_2NH_3)_2Cs_3Pb_4Br_{13}$, which presents higher stiffness[30], allowing us to study a different mechanical response range. Our findings show that a strain <1%[25–27] is sufficient to change the PL spectral features, with the emergence of additional PL peaks different from the bandgap emission at low-temperature. Indeed, we found an onset temperature of 230K, supporting the role of the thermal expansion coefficients mismatch between the flake and $SiO_2$ on the strain and confirming that an in-plane tension is generated during cooling. Furthermore, we found that the relative intensity of these PL peaks is sensitive to the local strain generated around the $SiO_2$ rings and to the size of the ring (height and diameter). By micro-Raman spectroscopy mapping and numerical mechanical simulation analysis, we demonstrate that these PL peaks result from the tensile and compressive deformation of the flake. Therefore, our work provides novel insight into the correlation between structural and optical properties in 2D HOIPs.

Moreover, we demonstrate how the selection of the material allows to detect small strain changes and the possibility to modulate the PL emission, interesting for sensing and optoelectronic devices applications.

We choose $(C_6H_5CH_2CH_2NH_3)_2(Cs)_3Pb_4Br_{13}$ 2D HOIP (n=4) for this study. This material shows higher stiffness than other 2D HOIPs used in previous studies (see **Table S1**, for Young's modulus and hardness values[25,29,30]). Moreover, as shown for lead iodide 2D HOIPs at room-temperature,[25] its n=4 structure makes the optical bandgap sensitive to the strain. Additionally, it contains phenethylammonium ($C_6H_5CH_2CH_2NH_3^+$ = $PEA^+$) and $Cs^+$ as cations, which enhance the ambient and moisture stability in optoelectronic devices.[33,34] **Figure 1**a shows a schematic of the crystal structure from n=1 to n=4.[35] Figure 1b shows the progressive redshift of the photoluminescence emission of the crystals at room-temperature from n=1 to 4,[35,36] confirming the formation of $(PEA)_2Cs_3Pb_4Br_{13}$ (see **Figure S1** for additional characterization). Moreover, temperature-dependent photoluminescence measurements show a systematic blueshift when temperature increases (Figure 1c, see **Figure S2**a-b for peak position, linewidth and integrated area evolution), similar to the 3D counterpart $CsPbBr_3$.[37] Additionally, to exclude phase transitions effect in the strain experiments, we checked by temperature-dependent micro-Raman spectroscopy the peak position and linewidth of the ~44 cm$^{-1}$ and ~70 cm$^{-1}$ vibrational modes ascribed to the distortion of the $[PbBr_6]^{4-}$ octahedra and translation of $Cs^+$, respectively[38,39]. The monotonous shift towards lower frequency of these Raman modes and their broadening as the temperature increases discard a phase transition in the temperature range under study (80-300K)[38,40] (Figure 1d, see Figure S2c-d for Raman spectra, peak position and linewidth evolution).

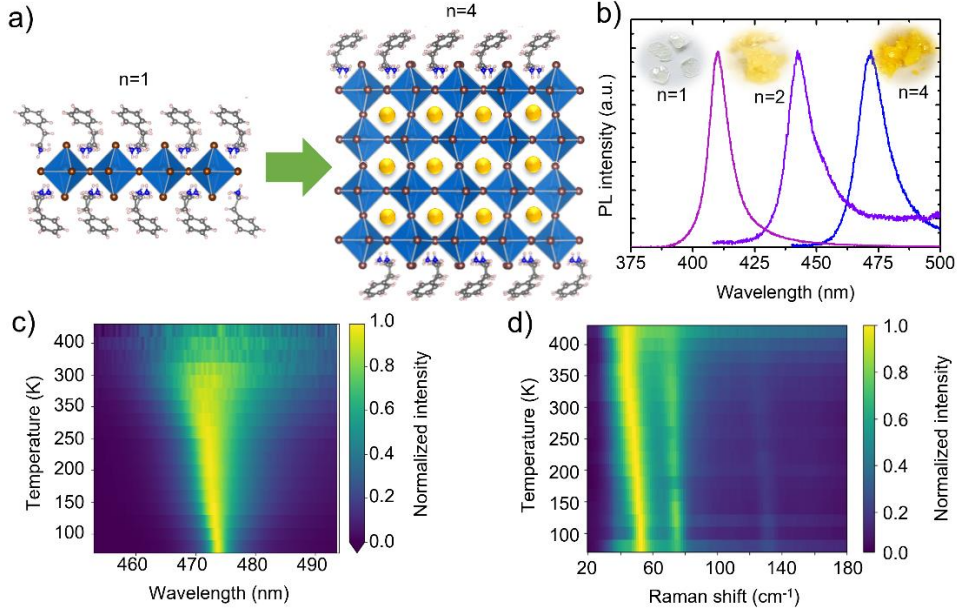

**Figure 1.** (a) Schematic representations of the crystal structure for $(PEA)_2PbBr_4$ (n = 1) and $(PEA)_2Cs_3Pb_4Br_{13}$ (n = 4), elaborated from crystal structures reported in refs [35,41]. (b) Photoluminescence spectra of the bulk single crystals of $(PEA)_2PbBr_4$ (n =1), $(PEA)_2CsPb_2Br_7$ (n = 2) and $(PEA)_2Cs_3Pb_4Br_{13}$ (n = 4). The insets show photographs of the synthesized crystals. (c, d) Representative temperature-dependent photoluminescence and Raman spectral mapping collected from a $(PEA)_2Cs_3Pb_4Br_{13}$ (n = 4) flake on $SiO_2/Si$ substrate.

To promote strain in the $(PEA)_2Cs_3Pb_4Br_{13}$ (n=4) flakes, we use $SiO_2$ rings, which provide bending points but also freedom to the suspended part of the flake to accommodate the strain. We fabricated arrays of 10×10 rings covering at least 50×100 μm by laser lithography and ion milling on 300 nm $SiO_2$/Si substrates (see Methods). The rings were separated from 25 to 50 μm center to center. We selected diameters (Ø) of 5, 10 and 15 μm; a width of 1 μm in all cases; and a height (H) of 100, 200 and 260 nm (see **Figure S3** for atomic force microscopy -AFM- analysis). Subsequently, we placed the mechanically exfoliated flakes on the $SiO_2$ rings by all-dry viscoelastic stamping[42] as illustrated in **Figure 2**a (see Methods). The flakes that could be successfully placed on the rings without breaking while keeping their mechanical stability for months were 60-600 nm thick, as shown in Figure 2b-c (see **Figure S4** for representative examples of each ring size and thickness). The AFM topography shows that the flakes do not remain flat, and that typically a segmental dome forms on top of the ring, as also confirmed by SEM (**Figure S5**). The formation of the dome is promoted by the elastic properties of the flake combined with the presence of the ring, which acts as support and induces flake bending during the stamping process. Indeed, the ring size affect the formation of the dome. A small height (100 nm) with large diameter (15 µm) leads to no dome formation since the flake touches the substrate in the center of the ring (Figure S4c). We observe excellent reproducibility across different samples (**Table S2**), indicating that an elastic process occurs. However, for simplicity, in the main text we focus our attention on the ring with H=200 nm and Ø=10 µm, as a representative case.

To compare the order of magnitude of the generated strain with other published works with HOIPs,[25,27] we estimated the strain ($\varepsilon$) considering the flake thickness ($h$ ~400 nm) and the radius of curvature ($R$ ~93.8 µm) as $\varepsilon = \frac{h}{2 \cdot R}$, where $R$ is estimated from the AFM images (90-100 µm) and determined by $R^2 = (R - \sigma)^2 + \left(\frac{\omega}{2}\right)^2$, $\sigma$ (~300 nm) and $\omega$ (~15 µm) being the height and diameter of the dome, respectively.[25,27] The estimation yields a strain of ~0.2% (achieving up to 0.4% - Table S2), indicating that we are in the range of other works (0-1.72%).[25,27]

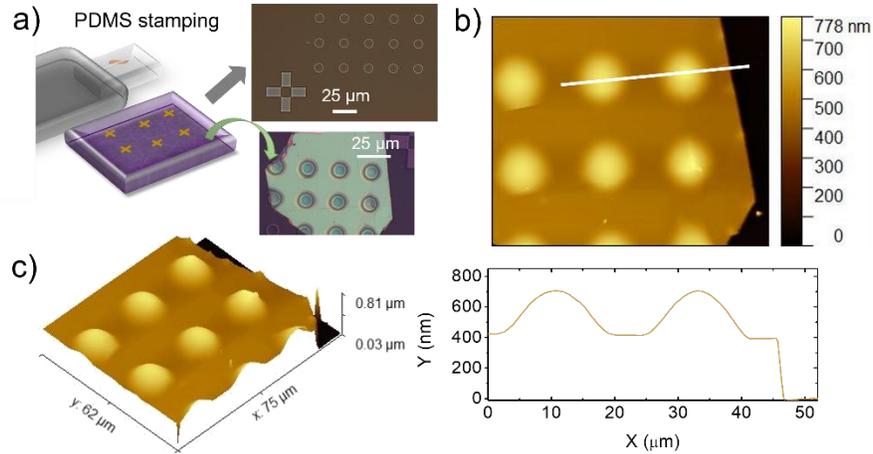

**Figure 2**. (a) Schematic representation of the stamping process used to transfer the flakes from the polydimethylsiloxane (PDMS) stamp onto the $SiO_2$ ring patterns (H = 200 nm, Ø = 10 µm) on the $SiO_2$/Si substrate. (b) Representative topography AFM image of a 400 nm thick flake on $SiO_2$ rings (H = 200 nm, Ø = 10 µm), whose optical microscope image is shown in (a). At the bottom the corresponding AFM profile is shown. (c) The corresponding 3D view of the same flake.

We evaluate the effect of the strain on the optical properties of the flake by temperature-dependent micro-photoluminescence spectroscopy. Considering the laser wavelength (405 nm) and the flake thickness, we collected the signal at least from the top 160 nm (see **Figure S6**). Results at room-temperature show slight changes in peak position (~1 nm blueshift) and broadening (~2 nm) (**Figure S7**), as expected, considering the strain involved[25]. However, at low-temperature (80K), the PL emission becomes narrower (FWHM from ~11 nm to ~3 nm) and several PL features can be resolved in the strained samples (**Figure 3**a, see **Figure S8** for strain-free flakes). The PL spectra of unstrained flakes (on a flat $SiO_2$/Si substrate) show a single PL peak ($P_{bg}$) at 474±1 nm (FWHM~3 nm) corresponding to the material's bandgap. In contrast, the strained flakes present a broader emission (FWHM~6 nm) consisting of three well resolved peaks at P1=472.0±0.5 nm (FWHM=2.6±0.5 nm), P2=474.0±0.5 nm (FWHM=1±0.1 nm) and P3=477.0±0.5 nm (FWHM=2.8±0.5 nm), whose peak position and FWHM vary slightly (~0.5 nm) across the flake, as expected considering the magnitude of the generated strain[25] (see **Figure S9** for fitting examples and PL spectra evolution). While P2 corresponds to the bandgap of the pristine perovskite, P1 is blueshifted 2 nm and P3 is redshifted 3 nm from the bandgap. These additional peaks can arise from tensile and compressive strain, which promote a widening or narrowing of the bandgap, respectively, as demonstrated in similar 2D HOIPs.[21,24,25,29,43] The relative intensity of P1 and P3 vs. P2 increases near the edge of the ring (Figure 3b-c), although it is not constant along the circumference of the ring. The fact that the ratio is not constant throughout the sample evidences that there are other secondary effects influencing the PL emission. One is likely interlayer sliding without octahedral tilting, already observed in other 2D HOIPs under mechanical force,[8,44] and here presumably caused by the stamping process. We cannot detect this effect from our AFM measurements, but it is expected to be inhomogeneous and to vary from sample to sample. We note that the three PL peaks appear throughout the strained flake, indicating that the whole flake (and its layers) accommodates the strain. Most importantly, new features emerge below 230K from the temperature-dependent analysis of the three PL peaks (**Figures S10-27**). Moreover, we observe a progressive decrease of the P1/P2 and P3/P2 ratio while increasing the temperature. This points to the presence of strain arising from the different thermal expansion coefficients ($\alpha$) of the flake and SiO$_2$. Since $\alpha_{SiO2}<\alpha_{HOIP}$, an in-plane tension is generated during cooling similarly to thin films.[45,46] The effect of the thermal expansion on our domes is also influenced by the geometry, as the flake is supported at the ring but free-standing in its vicinity. Changing the ring height from 200 to 100 nm we observe differences in the PL maps. At 100 nm height, the flake conforms to the ring in such a way that the strain is larger in the center, where the P1/P2 and P3/P2 ratios are higher in intensity. Combining H=100 nm and Ø=15 μm, there is no dome formation, the flake gets wrinkled on the ring, and the P1/P2 and P3/P2 ratios are larger in that area (Figure S12). When increasing the height to 200 or 260 nm, a larger lateral force is needed to form and support the dome, leading to larger strain at the edges, reflected by higher P1/P2 and P3/P2 ratios in that area. For H≥200 nm, the increase of the ring diameter has less influence, and higher P1/P2 and P3/P2 ratios are observed at the same area corresponding to the ring edge.

Temperature-dependent micro-Raman spectroscopy provides insight into the kind of strain generated and its consequences at the structural level. We collect signal from the whole thickness of the flake using a 532 nm laser (Figure S6), and monitor the vibrational modes of the perovskite[38,39] at ~48 cm$^{-1}$ and ~72 cm$^{-1}$ (Figure 3c), to evaluate their role in the mechanical relaxation. At 80K, we observed a systematic shift towards lower frequencies without significant changes in the linewidth in both Raman modes in the SiO$_2$ ring edge where the PL already highlighted the strongest changes. Although there is not a perfect match between the PL and Raman maps, there is a correlation: the larger the P1/P2 ratio corresponds to regions of redshift of the R1 and R2 Raman modes, as shown in Figure 3c. Furthermore, the same temperature onset seen in the PL spectra (~230K) is observed in the Raman maps (Figures S10-27). Experimentally, differences in the PL and Raman mapping can be attributed to strain-related factors affecting the local band structure, such as interlayer sliding mentioned above, to which PL emission is more sensitive than Raman spectroscopy.[8,21,22,44,47] Despite these factors, the observed systematic shift in the micro-Raman mapping in the 80-230K range indicates that the system responds to the generated strain with the deformation of the inorganic layers rather than the organic layers: the system adjusts the [PbBr$_6$]$^{4-}$ octahedral tilting and/or relaxes the crystal lattice, while the octahedral interlayer spacing remains almost unchanged.[21,24,25] Indeed PEA$^+$ molecules, are larger in volume than alkyl chains and quite rigid due to the $\pi - \pi$ interaction of their aromatic rings, minimizing the possibility of interlayer deformation[30,40,48,49] (see Table S1 for the high values of Young's modulus and hardness of 2D HOIPs with PEA$^+$). Since both Raman modes redshift, we may conclude that the generated strain corresponds to a tensile strain,[50] which modifies the $[PbBr_6]^{4-}$ octahedral tilting[21,25,29,43] and/or causes a small lattice expansion[24]. These changes in the crystal structure lead to a widening of the bandgap and to the appearance of PL peak P1 at 472 nm blueshifted vs the initial bandgap. In contrast, the less intense P3 observed at 477 nm may correspond to compressive strain[21,25,29,43] or to defects and self-trapped excitons[47] promoted during the strain accommodation. The P2 remains at 474 nm, corresponding to the bandgap emission, suggesting that there are unstrained regions on the flakes. To evaluate the influence of defects and self-trapped excitons[47] on the P1/P2 and P3/P2 intensity ratios, we carried out power-dependence experiments at 80K (**Figures S28-29**). Since the ratios do not depend on the power excitation, defects do not affect the PL emission.[51,52]

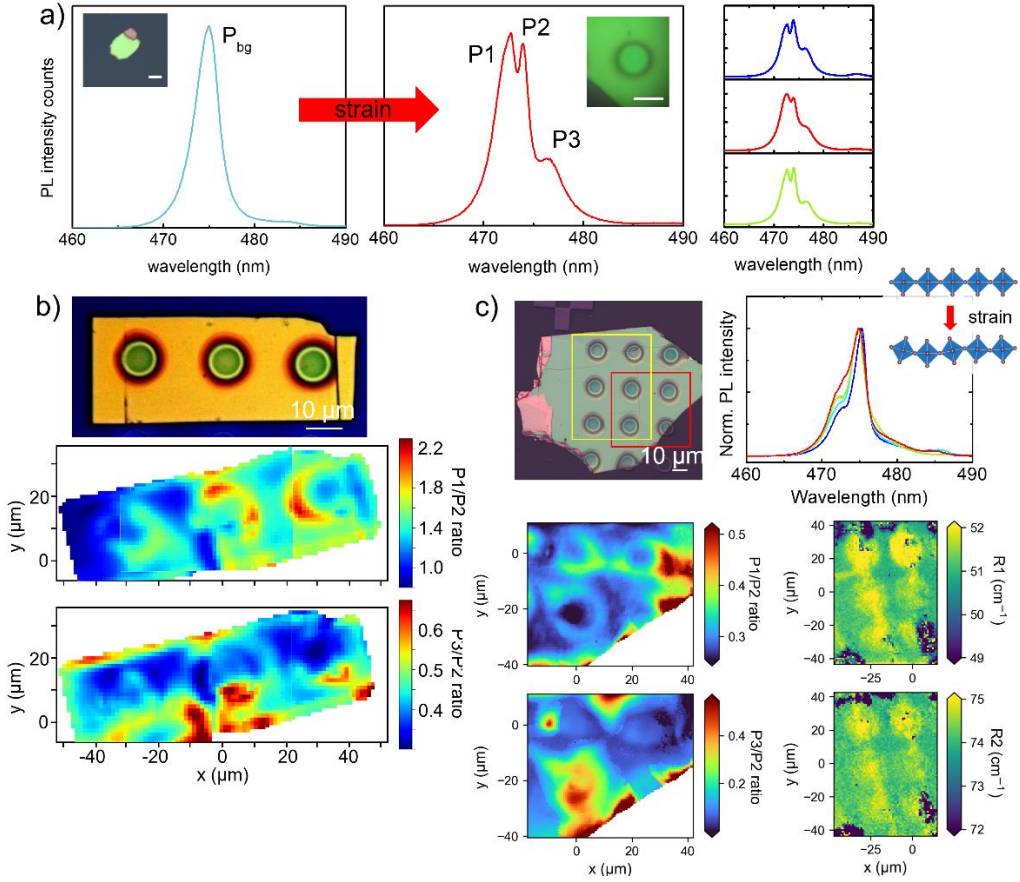

**Figure 3.** (a) Representative photoluminescence spectra at 80K for a flake without strain (blue line) and a flake under strain on a SiO$_2$ ring (H = 200 nm, Ø = 10 μm) (red line) together with three PL spectra taken from (b) in different regions, keeping the mapping color code. Scale bars in the insets correspond to 20 μm (unstrained) and 10 μm (ring sample). (b) Optical microscope image of a flake on a SiO$_2$ ring (H = 200 nm, Ø = 10 μm) and P1/P2 and P3/P2 intensity ratio maps at 80K showing an increase of the P1 and P3 emission intensity vs P2 around the SiO$_2$ ring. (c) Left: Optical microscope image of a flake on a SiO$_2$ ring (H = 200 nm, Ø = 10 μm) and P1/P2 and P3/P2 intensity ratio maps corresponding to the red square, showing an increase of the P1 and P3 emission intensity vs P2 around the SiO$_2$ ring. Right: R1 and R2 Raman modes (position) maps corresponding to the yellow box are also displayed, showing a shift towards lower frequency for both vibrations around the SiO$_2$ ring. PL and Raman spectra collected at 80K. The inset shows a schematic illustration of octahedral tilting variation under strain.

These results are supported by an analysis of the strain distribution inside the flake after being placed on the ring. We use finite-element simulations[53] at low- (**Figure 4**) and room-temperature (**Figures S30-31**) based on the topography obtained from AFM measurements for flakes on SiO$_2$ rings (H=200 nm, Ø=10 µm) and the reported elastic parameters (Table S1 and Methods). When the flake is placed on the ring, it is subject to biaxial strain (Figure 4b-c). From these simulations, we observe that independently of the temperature the range of strain is similar (-0.6 to 0.6%) and the net strain in the vertical direction is zero. Specifically, we find that in the internal region of the ring the maximum tensile strain on the top of the flake corresponds to a maximum of compressive strain on the bottom. Moreover, at the edges of the ring the strain is opposite, and the top of the flake suffers from compressive strain while the bottom undergoes to tensile strain. The co-existence of tensile and compressive strain supports the appearance of the strain-related PL peaks in the whole strained flake. The PL emission at 472 nm (P1) was already ascribed to tensile strain from the Raman results. Moreover, the PL emission at 477 nm (P3) can be ascribed to the narrowing of the bandgap caused by the compressive strain present in the flake.[21,25,29,43] However, the simulations are unable to clarify the scenario observed at 80K by micro-PL and Raman mapping in the region around the rings (Figure 3b-c). In fact, the simulations are designed to find the equilibrium of internal forces in the system and performed assuming the flake as a uniform and isotropic material. Simulations along the z-direction of the strain profile considering the hybrid nature of the flake show a kink when arriving at the molecular layer, indicating that the anisotropy affects the distribution of the strain, although the range of the strain is not affected (**Figure S32**).

We observe large differences between PL and Raman spectroscopy at low temperature (T<230K) but not at room-temperature. As mentioned above, a plausible explanation for this is the presence of thermally induced strain[45,46] (not addressed in our simulation model), due to the different thermal expansion coefficients of the flake and $SiO_2$. Additionally, regarding the effect of the thermal expansion on our domes, the point where the flake is in contact with the $SiO_2$ substrate close to the ring is relevant. We note that the inversion of the strain (tensile on the bottom, compressive on the top) occurs near this support point, favoring additional deformation of the flake during cooling. Therefore, this thermal strain can account for the systematic redshift in the Raman modes and for the enhancement of the P1/P2 and P3/P2 ratios in that region. In contrast, in the unstrained flakes the tension due to the thermal expansion is equally distributed along the flake and any changes remain hidden by the temperature-dependent expected shifts in PL and Raman spectra.

Overall, our results indicate that the effect of strain on the PL emission of 2D HOIP flakes cannot be explained with simple mechanical arguments. The emergence of additional PL peaks at 80K throughout the strained flakes clearly shows the sensitivity of the band structure to strain, with the coexistence of peaks related to tensile (P1) and compressive (P3) strain, while maintaining the unstrained bandgap emission (P2). By contrast, Raman spectroscopy cannot discriminate the strain distributions over the whole flake but shows the presence of tensile strain at ring edges. In the same region, the PL peak originating from tensile strain is stronger. Correspondingly, numerical simulations show the co-existence of tensile and compressive strain over the dome surfaces, with a gradient along the flake thickness. The predominance of the tensile strain peak in the PL suggests that, upon cooling from 230K to 80K, a differential thermal expansion brings the microstructure to an equilibrium configuration that is not fully captured by the Raman mapping, nor by the simulation.

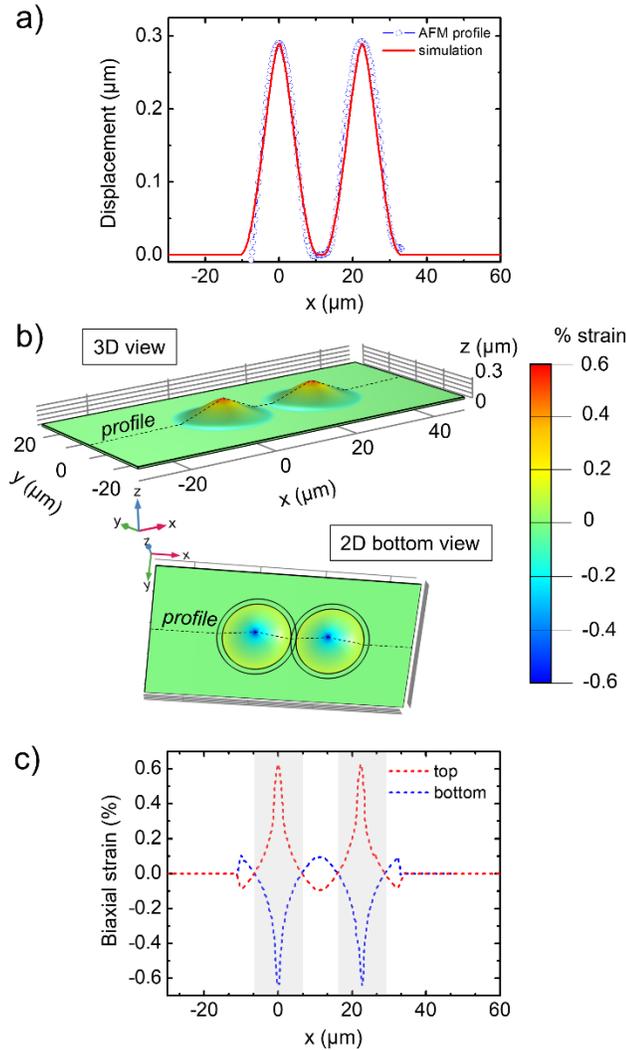

**Figure 4**. (a) Comparison between the AFM profile of two nearby domes collected at room-temperature and the corresponding finite-element simulation carried out at 80K. (b) The corresponding 3D and 2D bottom view of the mechanical simulation results at 80K of a prototypical flake (400 nm thickness) on a $SiO_2$ ring (H = 200 nm, Ø = 10 µm). The color scale shows the distribution of the strain: red (tensile) and blue (compressive) in the color scale bar. The concentric circles shown in the bottom view are the anchoring points for the mechanical simulation. (c) The corresponding strain profiles taken from (b) and indicated as dashed lines in the images. The shaded area corresponds to the area of the $SiO_2$ rings.

In conclusion, we investigated the temperature-dependent photoluminescence response of 2D $(PEA)_2Cs_3Pb_4Br_{13}$ flakes to biaxial strain on $SiO_2$ rings. The $(PEA)_2Cs_3Pb_4Br_{13}$ mechanical properties allowed us to optically detect a strain <1%. We found that it is possible to discern small changes (~2 nm) in the optical bandgap promoted by strain using low-temperature (80K) photoluminescence measurements. Additionally, combining temperature-dependent micro-Raman spectroscopy and reverse engineering strain modeling, our work elucidates the nature of the bandgap modulation with the mechanical and thermal strain as result of changes in the inorganic $[PbBr_6]^{4-}$ octahedral tilting and/or lattice expansion and how the whole flake accommodates the generated strain. These findings provide new insight into the relevance of the material design and the choice of the platform used to induce strain in a device-compatible substrate, as well as the relevance of the combination of simulations and different spectroscopic techniques to understand and develop new strain-based optoelectronic and sensing applications using 2D HOIPs.

ASSOCIATED CONTENT

**Supporting Information**. Experimental and computational methods; Mechanical properties of layered hybrid organic-inorganic metal-halide perovskites; Additional Raman spectroscopy characterization; Characterization of the $SiO_2$ rings and flake/ring structures by SEM and AFM; z-depth penetration of the lasers; Supplementary photoluminescence and Raman spectroscopy data; Evolution of the temperature-dependent photoluminescence and Raman maps with the $SiO_2$ ring dimensions; Power-dependence of the photoluminescence maps; and Additional finite element simulations.

The following files are available free of charge.

Supporting Information (PDF)

AUTHOR INFORMATION

**Corresponding Author**

*Beatriz Martín-García (b.martingarcia@nanogune.eu)

**Author Contributions**

The manuscript was written through contributions of all authors. All authors have given approval to the final version of the manuscript.

**Funding Sources**

This work is supported by the Spanish MICINN under Project PID2019-108153GA-I00 and under the María de Maeztu Units of Excellence Programme (MDM-2016-0618).

ACKNOWLEDGMENT

This work is supported by the Spanish MICINN under Project PID2019-108153GA-I00 and under the María de Maeztu Units of Excellence Programme (MDM-2016-0618). B.M-G. thanks Gipuzkoa Council (Spain) in the frame of Gipuzkoa Fellows Program and to Prof. A. Mateo-Alonso (Molecular and Supramolecular Materials Group - POLYMAT) for the access to the Chemistry Lab to grow the crystals. M.G. acknowledges support from la Caixa Foundation (ID 0010434) for a Junior Leader fellowship (Grant No. LCF/BQ/PI19/11690017). Authors thanks Dr. E. Goiri Little for reading and revising the manuscript.

ABBREVIATIONS

AFM, atomic force microscopy; HOIP, hybrid organic-inorganic metal halide perovskite; PEA, phenethylammonium; PL, photoluminescence; SEM, scanning electron microscopy.